\documentclass[a4paper]{ESASPCS13Style}
\usepackage{epsfig}
\def\lesssim{\mathrel{\hbox{\rlap{\hbox{\lower4pt\hbox{$\sim$}}}\hbox{$<$}}}}
\def\gtrsim{\mathrel{\hbox{\rlap{\hbox{\lower4pt\hbox{$\sim$}}}\hbox{$>$}}}}

\begin{document}

\title{Current views on low-mass star formation}

\author{Francesco Palla and Sofia Randich} 
\institute{INAF--Osservatorio Astrofisico di Arcetri, Largo E. Fermi, 5,
50125 Firenze, Italy}

\maketitle 

\begin{abstract}

The process that leads to the formation and early evolution of low-mass
stars is in a broad sense well understood theoretically and carefully traced
observationally. The largest uncertainties in this framework reside in the
poorly known initial conditions for the origin and gravitational collapse
of dense cores within molecular clouds. In this review, we will discuss
the evidence for an accelerated pattern of star formation in clusters
and associations, the physical origin of this phenomenon, and the age spread
in low-mass star forming regions. On the latter issue, we will show how the
observation of lithium depletion in young clusters, such as the Orion Nebula
Cluster and the Upper Scorpius association, can provide an independent
estimate of their ages and age spreads with important implications on the 
understanding of the initial conditions and history of star formation.

\end{abstract}

\section{Introduction}
  
Boldly stated, the formation of low-mass stars (solar and below) is a rather
well understood process, both theoretically and observationally. Empirically,
circumstantial evidence has been accumulated in the last three decades over
the whole e.m. spectrum, from X-rays to the millimeter domain. This progress
has led to a paradigm that allows to follow the evolutionary path of a
nascent low-mass object from the initial seed, a dense molecular core within
more diffuse gas, to the formation of an accreting protostellar core with its
attendant outflow, to the final buildup through a circumstellar disk (e.g.,
Shu et al. 1999). Each step is characterized by a complex array of physical
processes that have been theoretically laid out and analyzed in detail with
semianalytic techniques or by means of sophisticated numerical simulations.
Of course, the extremely wide range of physical scales involved prohibits (or
limits substantially) the ability to follow in a single model the
fragmentation, collapse and accretion/ejection phase of a low-mass object
including all the essential processes.  As a result, some fundamental
questions are still only partially answered, including the origin of
outflows, the mechanisms responsible for mass accretion in disks, etc.

Another important aspect that has emerged from observational studies is that
most stars in the Galaxy form in clusters and associations and that the
occurrence of isolated objects and small groups represents a rare phenomenon
restricted to specific regions, such as Bok globules and bright rimmed
clouds.  Considering the solar vicinity, the vast majority of the young
stellar population within 0.5 kpc from the Sun is contained in a variety of
T, R, and OB associations (about 30 known groups), plus 15 embedded clusters
of varying richness (e.g., Lada \& Lada 2003), and a number of relatively
sparse moving groups and associations (e.g., Lepine \& Gregorio-Hetem 2003).
Thus, in order to discuss the problem of low-mass star formation in a more
realistic way, one should be able to answer the basic question of how do
stars form in groups, clusters, and associations. Equally important is to
consider the relation of the young stars found in these systems to the
molecular clouds and cloud complexes that spawn them.

The problem of the origin of low-mass stars is obviously directly linked to the
question of what supports a massive cloud against its self-gravity {\it
before} the production of an interior stellar group: how do individual dense
cores, each capable of forming single and binary stars, arise within the
magnetized and turbulent medium that characterizes molecular clouds?
Considering the main interests of the participants at the {\it Cool Stars}
meetings, in this review we will focus the attention on these global issues
rather than the specific problems related to the formation of a single/binary
system. As we will see, there is a lot of fertilization in the field of star
formation that can come from studies traditionally carried out by the stellar
community.

\section{Models of global star formation} 
\label{sec:modes}

Answering the above questions requires a broader understanding that is
currently at hand, one that connects the birth of individual stars to the
growth and evolution of clouds on a multiparsec scale. A big effort in this
respect is represented by a variety of numerical simulations of cloud
evolution and star formation based on gravoturbulent (e.g., MacLow \& Klessen
2004) or MHD turbulent (e.g., Padoan et al. 2004) models.  In simple terms,
the picture that emerges from these models is one in which molecular clouds
are formed in large-scale turbulent HI flows and that the intersection of
these flows compresses the gas to the point that it collapses, fragmenting
simultaneously into stars over a broad front.  Star formation is thus
considered as a process that occurs promptly in localized regions where
turbulence is dissipated in a dynamical timescale (Elmegreen 2000). Then,
molecular clouds can sustain star birth only for few million years.

This {\it dynamic} picture contrasts sharply with that suggested by, e.g.,
Palla \& Stahler (2002)  in which the
formation of dense cores is not seen as a random event in the cloud medium,
but is supposed to occur in response to some global change in that medium.
One such change would be large-scale, gravitational contraction of the parent
cloud.  This contraction presumably occurs through gradual loss of the
mechanical support driven by ambipolar diffusion, {\it i.e.} the clouds
evolve in a {\it quasi-static} fashion (see also Tassis \& Mouschovias 2004).
The time scale for this process is of order $\sim$10~Myr and exceeds
that predicted by the dynamic models.

Empirical evidence in favor of the slow mode of star formation comes from the
reconstruction of the star formation history in clusters and associations
using the classical method of placing stars in the HR diagram to infer
isochronal ages.  The application of this method to the most conspicuous
regions in the solar vicinity with a statistically significant population of
young stars has led to the following conclusions (Palla \& Stahler
2000).  In general, it is found star formation began at modest levels roughly
$1\times 10^7$~yr in the past and increased rapidly toward the present epoch
with an {\it acceleration} pattern. This trend is similar in different
systems, although the time scale of the acceleration varies from region to
region, typically between 1 and 3$\times 10^6$ yr. The fast acceleration must
be followed by a prompt deceleration to limit the global star formation
efficiency of each unit to the observed low values (few percent). Evidence
for such a steep decline is seen in regions such as Upper Scorpius and
$\lambda$ Ori where the activity reached a peak $\sim 3\times 10^6$ yr ago
and then dropped essentially to zero owing to the efficient removal of the
dense gas by the radiative and mechanical effects of massive stars.

The physical interpretation of this empirical pattern is that the formation
and evolution of dense cores and stars therein occurs in response to a
global, quasi-static contraction of the parent cloud. Then, the formation of
stars is seen as a {\it threshold} phenomenon where a minimum column density
of molecular gas must be reached for dense core formation which then produce
stars on a relatively brief time scale.  Although these findings seem to be
robust, a number of important effects (incompleteness of the knowledge of
both the dense gas and stellar population in each system, uncertainties in
PMS tracks and isochrones and in stellar parameters, etc.) should be
considered carefully in each case.  The limitations of the method have been
analyzed by Hartmann (2002, 2003), while additional discussion is provided by
Palla (2004). Let me conclude this section by noting that the two models
discussed above have a strong similarity, namely the accelerating character
of star formation, but depart in an essential respect:  in the dynamic mode
there is no past activity and star formation occurs in a single major burst,
while in the quasi-static mode there is a long history that can be traced
back to its beginnings.

\section{Age spreads in star forming regions}
\label{sec:spread}

The presence of a relatively old population in young clusters and associations
seems to indicate that molecular clouds can produce stars over extended
periods ot time.  Whether the inferred {\it age spread} is real or not is the
central question to be answered. Although much information is available from
the study of regions of ongoing star formation, a definite answer cannot come
from them since they are still actively turning dense gas into stars and we
cannot infer how long this process will continue in the future. Thus,
active SFRs can only provide a partial estimate of the true age spread. This
simple consideration rules out quite a large number of well known complexes,
including Taurus-Auriga, Chamaeleon, and $\rho$ Ophiuchi in which the amount
of available gas for future episodes of star formation greatly exceeds that
already converted into stars.

In contrast, the ideal laboratories to derive reliable age spreads are
represented by two other classes of objects: fully exposed, young clusters
and associations ($t<$10~Myr), and intermediate-age ($t\sim$10--30 Myr) open
clusters. In both cases, the key to determine their age and age spread is the
use of the {\it lithium depletion history and boundary} tests (LDB) as a
powerful clock (e.g., Basri et al. 1996). This method is well known to the
stellar community for its successful application to a number of older open
clusters, such as $\alpha$ Per, IC 2391, NGC 2547, and the Pleiades, with
significant revisions of their ages (e.g., Jeffries 2004). In the remaining
part of the section, We will discuss the potential of its application to
younger systems.

\subsection{The lithium test}
\label{sec:lithium}

The lithium test rests
on the ability of young stars to deplete their initial lithium content during the early
phases of pre--main-sequence (PMS) contraction. Young, low-mass mass stars ($M_\ast \lesssim$1~M$_\odot$) begin their PMS
phase with the full interstellar supply of lithium since the central regions
are too cold to ignite nuclear burning during the protostellar phase.  As
contraction proceeds, the critical temperature for lithium reactions ($\sim
2.5\times 10^6$~K) is reached, and the initial content is readily depleted in
fully convective, sub-solar stars (M$\lesssim$0.5 M$_\odot$).  It has been
shown that the physics required to study the depletion history as a function
of age has little uncertainty, since it depends only on the stellar mass and
radius (e.g.  Bildsten et al. 1997). Accordingly, fully convective stars in
the range 0.2--0.5 M$_\odot$ start to deplete lithium after about 2~Myr, and
completely destroy it in approximately $\sim$10~Myr (e.g., Baraffe et al.
1998). Lower mass stars take much longer to burn lithium, and there is a
sharp transition (i.e., {\it the boundary}) between fully depleted objects
and those with the initial lithium content.  The mass at which the boundary
occurs reveals the age of the cluster.  The predicted region of partial and
full lithium depletion is sketched in the HR diagram of Figure 1. Stars more
massive than $\sim$0.8 M$_\odot$ only burn a small fraction of the
interstellar value; on the other hand, fully convective objects
(M$\lesssim$0.5 M$_\odot$) readily consumme lithium while contracting toward
the ZAMS.

\begin{figure}[ht]
\begin{center}
\epsfig{file=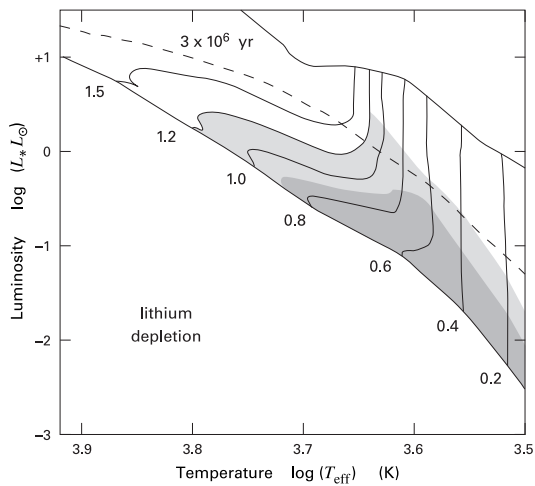, width=7cm}
\end{center}
\caption{The region of lithum burning and depletion in the HR diagram.
The light shaded area is for a depletion down to 1/10 of the interstellar
value, while the darker shading indicates larger depletion. Theoretical
tracks are from Palla \& Stahler (1999), while the values of lithium
depletion are from the models of Siess et al. (2000). The isochrone for 3 Myr
is shown by the dashed line.\label{fig1}}
\end{figure}

Returning to the criteria for selecting suitable astronomical objects,
the choice of open clusters with ages between 10 and 30 Myr is dictated by at
least three reasons. First, by this time the star formation activity is
certainly completed. Second, the age spread can be better estimated than for
older systems since isochrones at this age do not suffer from crowding.
Third, the LDB falls in the stellar regime (between 0.2--0.4 M$_\odot$ for
typical distances) and thus can provide strong constraints on their internal
structure. So
far, the LDB test has been determined in older systems where it falls in the
substellar regime.  The identification of the best systems to study is not
easy, due to an intrinsic paucity of open clusters of the appropriate age in
the solar vicinity and an observational bias introduced by the current strong
interest on the discovery and characterization of brown dwarfs in older
clusters rather than on the accurate measurements of lithium abundances in
low-mass stars.  However, it is important that a large effort be dedicated to
the study of younger clusters and the initial steps in this direction have
already been started with the completion of the CFHT Key Project on Young
Clusters (Bouvier et al. 2004, in preparation).

The other class of fully exposed, young clusters and associations may be
considered at first sight a surprising choice since one would not expect to
see any feature related to lithium burning and depletion.  However, we will
show below that this is not true in general and present the case of two
extremely interesting objects, the Orion Nebula Cluster (ONC) and the Upper
Scorpius Association (USA).

\subsection{Li-depletion in the ONC}
\label{sec:ONC}

The ONC is the best known star forming region in the solar neighborhood. It
is close ($d=450$~pc), away from the Galactic plane ($b\sim -$19$^\circ$),
and it lies in front of a giant molecular cloud, whose total extinction (up
to $A_V=50-100$~mag on the close edge) eliminates most of the background
confusion.  The richness of the stellar cluster ($n_\ast \sim 3500$~stars)
and density ($\rho_\ast \sim 2\times 10^4$~stars pc$^{-3}$ at the core) make
the ONC an ideal, and unique target for the study of the full mass spectrum
of young stars from 45 M$_\odot$ to less than $\sim$0.02 M$_\odot$ (e.g.,
Hillenbrand 1997, Slesnick et al. 2004). The reconstruction of the star
formation history indicates that the period of most active formation is
confined to a few $\times 10^6$~yr, and has recently ended with the dispersal
of the remnant molecular gas (Palla \& Stahler 2000).

\begin{figure}[ht]
\begin{center}
\epsfig{file=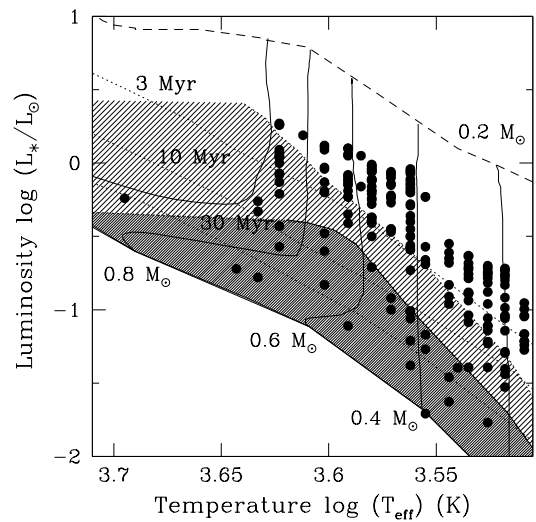, width=7cm}
\end{center}
\caption{HR diagram of the older population of low-mass stars of the ONC.
The hatched regions have the same meaning as in Fig.~1.
\label{fig2}}
\end{figure}

In addition to providing an average age of the ONC, the reconstruction of the
pattern of stellar births has revealed another important aspect regarding its
age spread:  namely, the existence of a small, but statistically
significant population of older stars with estimated ages in excess of
$\sim$5~Myr.  The distribution in the HR diagram of the low-mass members
(0.2--0.8 M$_\odot$) is shown in Figure 2. Here, we see that
there are more than about 80 stars that fall within the predicted
Li-depletion region.  Observations of a sample of $\sim$90 stars (membership
probability $>$80\%) with mass 0.4--0.8 M$_\odot$ and isochronal ages greater
than $\sim$1 Myr have been carried out using FLAMES$+$Giraffe on ESO-VLT2 and
the results are very encouraging (Palla et al. 2004, in preparation). As
shown in Figure 3, we find a decrease of the Li-abundance by a factor 5--10
in the coldest (T$\sim$3700~K) and faintest objects. Comparison with PMS
evolutionary models indicates that the observed Li-depletion corresponds to
stellar ages greater than $\sim$5 Myr.  These observations will be soon
extended to members of mass down to $\sim$0.2 M$_\odot$ with the goal of
probing the full depletion history.  Below $\sim$0.2 M$_\odot$, the LDB would
occur at an age $\gtrsim$30 Myr, well in excess of any realistic estimate of
the age spread of the ONC. Note that independent evidence for a population of
old stars in this cluster has been presented by Slesnick et al. (2004) in a
study of the spectroscopically confirmed brown dwarf population.  We conclude
that, despite its evident youth, the ONC is not simply the result of a single
burst of star formation. 

\begin{figure}[ht]
\begin{center}
\epsfig{file=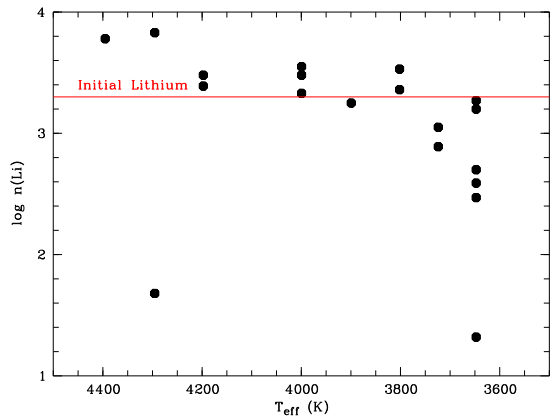, width=6cm, angle=-90}
\end{center}
\caption{Lithium abundances in stars of the ONC with mass $\simeq$0.4--0.5
M$_\odot$ (Palla et al. 2004, in preparation). 
\label{fig3}}
\end{figure}

\subsection{Li-depletion in the USA}
\label{sec:SCO}

The Upper Scorpius Association (USA) is the nearest OB
association to the Sun ($d$=145 pc) and the youngest of the three subgroups
that compose the Scorpius-Centaurus OB2 association. In the recent past, the
USA was a site of vigorous star formation activity that has produced a rich
group of objects over the full mass spectrum (de Zeeuw et al. 1999 for the
high mass and Preibisch \& Zinnecker 1999 for the extension to low masses).
Owing to the lack of dense molecular clouds in the vicinity (typically,
A$_V\lesssim 2$~mag) and of deeply embedded young stellar objects, one can
consider that the star formation activity in this region is basically
finished, thus allowing a full census of the stellar population  in the range
0.1--20 M$_\odot$ (Preibisch et al. 2002).

Using the isochronal method for gauging stellar ages and hence the star
formation history of the USA, Preibisch et al. (2002) find that the bulk of
the stellar population is characterized by a rather narrow age distribution,
centered at about 5 Myr ago. This property has been interpreted as evidence
for a short-lived episode of star formation, perhaps triggered externally by
the explosion of a nearby supernova (Preibisch \& Zinnecker 1999). However,
as in the case of the ONC, the USA displays a significant population of older
stars with estimated ages in excess of $\sim$5~Myr.  If true, this finding
suggests an alternative scenario to the rapid episode for the history of star
formation.

\begin{figure}[ht]
\begin{center}
\epsfig{file=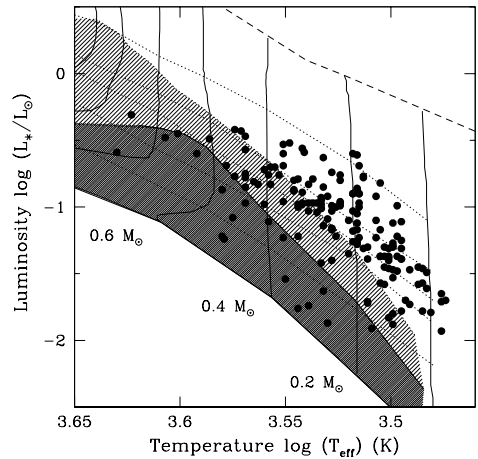, width=7cm}
\end{center}
\caption{HR diagram of the low-mass members of the Upper Scorpius Association 
(from Preibisch et al. 2002). The hatched regions have the same meaning as
in Fig. 1.
\label{fig4}}
\end{figure}

Interestingly, a large number of low-mass stars of the USA (about 70) fall in the
critical region where lithium depletion is expected to occur. Their
distribution in the HR diagram is shown in Figure 4. For all of them, the Li I
6708 \AA~line equivalent widths (EW) have been measured from low resolution
spectra (Prebisch et al. 2002). The exciting result is that most of the
objects for which substantial Li-depletion is expected indeed show lower
values of EW(Li). This is shown in Figure 5 where one can see that the oldest stars
of the sample display the largest decrease of EW(Li). Unfortunately, because
of the poor spectral resolution, Li-abundances could not be reliably measured
for comparison with the predictions of stellar models. The trend of Fig. 5
is quite intriguing since the stars with the lowest EW(Li) are also the least
massive objects of the sample for which the LDB is expected to show up.
Future high resolution spectroscopic observations should be carried out to
obtain accurate abundances and to test these initial results. In the meantime,
we can conclude that the history of star formation in the USA seems not to
be limited to a major single burst.

\begin{figure}[ht]
\begin{center}
\epsfig{file=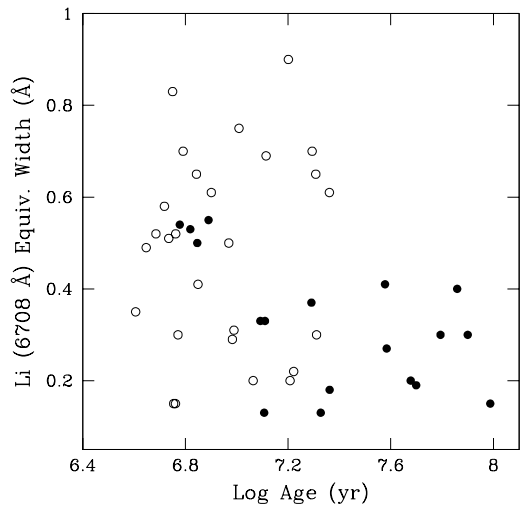, width=7cm}
\end{center}
\caption{Variation of the Li I 6708 \AA~ equivalent width with age for the 
low-mass sample of the USA that falls in the Li-depletion region of Fig. 4.
Filled circles correspond to stars in the dark region of the HR diagram of
Fig. 4, while empty circles are for those in the light grey area.
\label{fig5}}
\end{figure}

\section{Conclusion}
\label{sec:conclusion}

The initial tests on lithium depletion in the low-mass stars of two major
systems, Orion and Upper Scorpius, can offer a new way of attacking the
fundamental problem of determining the relevant time scale over which stars
form in molecular clouds.  Although the debate on the {\it rapid vs. slow}
mode of star formation is still open, the observational support for the
presence of an old population in young objects is becoming more compelling.
Additional measurements of lithium abundances in other nearby, young clusters
and associations can be (and will be) performed soon with existing
instrumentation. The hope is that the results will offer not only stronger
constraints on stellar interior models, but also more definite answers on the
history of low-mass star formation.

\begin{acknowledgements}{}
We are very grateful to Roberto Pallavicini who has generously allocated part
of the GTO time on FLAMES for the ONC observations and Ettore Flaccomio for
useful discussions.  Finally, it is a pleasure to thank the organizers of a
very interesting and informative meeting.  \end{acknowledgements}{}

\end{document}